\documentclass[12pt,a4paper,aps,nofootinbib]{article}
\usepackage{amsmath,amssymb,amsfonts,amsthm,bm,bbm,cancel,wasysym}
\usepackage{epsfig,graphics,graphicx,epstopdf,caption,subcaption}
\usepackage[numbers,sort&compress]{natbib}
\graphicspath{{Charts/}}
\usepackage{array,booktabs,colortbl,colordvi,multirow}
\usepackage{colordvi,color,xcolor}
\usepackage{hyperref}
\usepackage{rotating}
\usepackage{comment}
\usepackage{feynmf}

\usepackage[symbol]{footmisc}
\renewcommand{\thefootnote}{\fnsymbol{footnote}}

\begin{document}

\begin{center}

{\Large {\bf Revisiting the Hubble tension with an inverse power-law early dark energy}}\\

\vspace*{0.75cm}

{Quan Zhou and Sibo Zheng\footnote{Contact author: sibozheng.zju@gmail.com}}

\vspace{0.5cm}
{School of Physics, Chongqing University, Chongqing 401331, China}
\end{center}
\vspace{.5cm}

\begin{abstract}
\noindent

We propose a new early dark energy with a potential that has an  inverse power-law asymptotic tail to alleviate the Hubble tension. 
Fitting this model to the datasets of CMB+BAO+SN+H0DN,
one obtains a 68$\%$ confidence-level value of $H_0\sim 70.20^{+0.60}_{-0.47}$ km s$^{-1}$Mpc$^{-1}$ and a best-fit value of $H_0\sim 70.56$ km s$^{-1}$Mpc$^{-1}$ for index $n=10$.
We show that this model can be tested by future data of CMB and matter power spectra.
Finally, we briefly discuss fine-tuning problems related to this model.

\end{abstract}

\renewcommand{\thefootnote}{\arabic{footnote}}
\setcounter{footnote}{0}
\thispagestyle{empty}
\vfill
\newpage
\setcounter{page}{1}

\tableofcontents

\section{Introduction}
The Hubble tension refers to the discrepancy between the value of the Hubble constant inferred from the data of Cosmic Microwave Background (CMB) and that measured by the local experiments within the $\Lambda$CDM cosmology. 
Explicitly, the current Planck data implies a 68$\%$ confidence-level (CL) value of $H_{0}=67.36 \pm 0.54$ km s$^{-1}$Mpc$^{-1}$\cite{Planck:2018vyg},
while the SH0ES collaboration reported a value of $H_{0} =73.04 \pm 1.04$ km s$^{-1}$Mpc$^{-1}$ \cite{Riess:2021jrx} which is recently updated to $H_{0}= 73.17\pm 0.86$ km s$^{-1}$Mpc$^{-1}$ \cite{Breuval:2024lsv} through an improved distance ladder calibration, 
suggesting the discrepancy larger than $5\sigma$ order.
Consider that both the measurements are highly precise, 
the Hubble tension is unlikely to be caused by statistical fluctuations, and therefore motivates extensive investigations of new physics beyond $\Lambda$CDM. For reviews see e.g., \cite{DiValentino:2021izs, Abdalla:2022yfr,Perivolaropoulos:2021jda,CosmoVerseNetwork:2025alb}.

To date, early dark energy (EDE) \cite{Poulin:2023lkg} is one of the few scenarios that can reduce the tension below $3\sigma$.
More specifically, this EDE can be understood as a fundamental scalar degree of freedom beyond the Standard Model.\footnote{We refer the reader to  \cite{Poulin:2023lkg} for alternative EDEs models that lack particle interpretations and are studied through effective fluid approximation.}
In single-field realizations, there are several examples of EDE such as axion-like EDE \cite{Poulin:2018cxd, Smith:2019ihp}, Rock `n' Roll EDE \cite{Agrawal:2019lmo}, $\alpha$-attractor EDE \cite{Braglia:2020bym} and expotential EDE \cite{Sohail:2024oki},
among which the axion-like EDE offers the largest value of $H_{0}$,  
of order $\sim 70-72$ km s$^{-1}$Mpc$^{-1}$ depending on the datasets used. 
In multi-field realizations, the representative examples include new EDE \cite{Niedermann:2019olb, Niedermann:2020dwg, Niedermann:2021vgd, Niedermann:2021ijp,Cruz:2023cxy,Garny:2024ums,Garny:2025kqj}, multiple-axion EDE \cite{Bella:2026zuk} and decaying ultralight scalar \cite{Gonzalez:2020fdy}, of which the second case can give even larger value of $H_{0}\sim 72-73$ km s$^{-1}$Mpc$^{-1}$.

However, the aforementioned EDE models suffer from various fine-tuning issues, as summarized in \cite{Poulin:2023lkg}.
Regarding the axion-like EDE with index $n\geq 3$,  a construction of its ultraviolet theory is challenging \cite{McDonough:2022pku,Rudelius:2022gyu}.
For the new EDE models, their potentials are not protected by symmetries as that of the axion-like EDE, 
which may be spoiled by either renormalizable terms \cite{Poulin:2018dzj} or  radiative corrections.

Motivated by these problems, 
in this study we propose a new toy model of EDE with a potential that has an  inverse power-law asymptotic tail,
which differs from those of the existing toy models but alleviates the Hubble tension similarly. 
The rest of this paper is organized as follows. In Sec.~\ref{model}, 
we introduce the EDE model and describe its background and linear perturbation evolution. 
Sec.~\ref{MCMC} is devoted to present the datasets considered in this work, numerical methodology, and cosmological constraints,
where the goodness of fit will be measured by two different criteria.
In Sec.~\ref{signals}, we discuss the predicted signals of our EDE model in the CMB and matter power spectra. 
We conclude in Sec.~\ref{con}. 
The shooting algorithm used to map the EDE model parameters to phenomenological parameters is described in Appendix~\ref{sa}.

\section{The EDE model}
\label{model}
We consider the EDE described by the canonical scalar field $\phi$. 
At the homogeneous level, the background dynamics of $\phi$ is governed by the Klein-Gordon equation
\begin{equation}
	\ddot{\phi}+3H\dot{\phi}+\frac{dV(\phi)}{d\phi}=0 ,
	\label{eq:KG}
\end{equation}
where dots denote derivatives with respect to cosmic time. 
The energy density and pressure of the scalar field are given by
\begin{eqnarray}
	\rho_\phi &=& \frac{1}{2}\dot{\phi}^2+V(\phi),\\
	P_\phi &=& \frac{1}{2}\dot{\phi}^2-V(\phi),
\end{eqnarray}
in terms of which the equation-of-state parameter is given by
\begin{equation}
	w_\phi =
	\frac{\frac{1}{2}\dot{\phi}^2-V(\phi)}
	{\frac{1}{2}\dot{\phi}^2+V(\phi)} .
\end{equation}
Specifically, we choose the scalar field potential as
\begin{equation}
	V(\phi) =V_0 \left[1+\left(\frac{\phi}{f}\right)^2\right]^{-n},
	\label{eq:potential}
\end{equation}
where $V_0$ sets the characteristic energy scale of the potential, 
$f$ controls the field range, 
and $n$ determines the steepness of the potential away from the central plateau. 
For $n>0$, the potential is maximal at $\phi=0$ and decreases as $|\phi|$ increases. 
Due to the Hubble friction term in eq.(\ref{eq:KG}), 
the field is initially frozen at $\phi\sim 0$, resulting in a constant value of $V$. 
Once the expansion rate drops below a certain critical value, $\phi$ begins to roll down towards the asymptotic region where $V$ approaches zero.

\begin{figure}
	\centering
	\includegraphics[width=15cm, height=18cm]{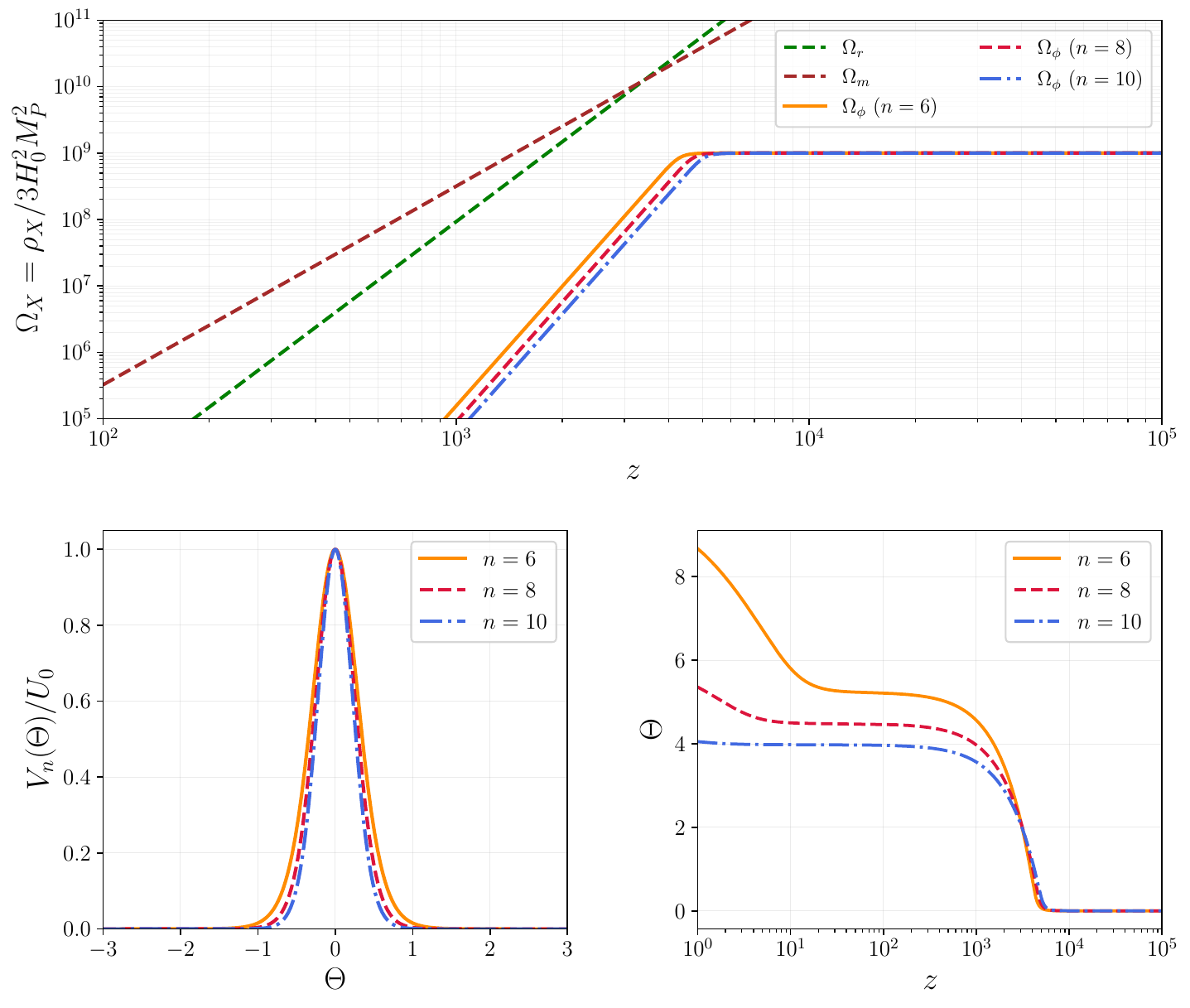}
	\centering
	\caption{An illustration of the evolutions of $\rho_\phi (z)$ (top), $V_{n}(\Theta)$ (bottom left) and $\Theta (z)$ (bottom right) for three values of $n=\{6,8,10\}$ with $U_0 = 3 \times 10^9$, $\alpha = 0.05$, and $\Theta_i = 10^{-3}$.}
	\label{background}
\end{figure}

\subsection{Background evolution}
The background expansion is determined by the Friedmann equation
\begin{equation}
	H=H_0 \sqrt{\Omega_m(z)+\Omega_r(z)+\Omega_\Lambda+\Omega_\phi(z)}=H_0 E(z),
	\label{Friedmann}
\end{equation}
where $H_0$ is the present-day value of Hubble parameter.
In order to solve eq.(\ref{eq:KG}) numerically, 
we introduce dimensionless variables
\begin{equation}
	\Theta \equiv \frac{\phi}{f}, \qquad
	\alpha \equiv \frac{f}{M_P}, \qquad
	x \equiv H_0 t, \qquad
	U_0 \equiv \frac{V_0}{H_0^2 M_P^2},
\end{equation}
where $M_P=(8\pi G)^{-1/2}$ is the reduced Planck mass. In terms of these variables, 
eq.(\ref{eq:KG}) is rewritten as 
\begin{equation}
\tilde{\tilde{\Theta}}= -3E\tilde{\Theta} - \alpha^{-2}\frac{dV_n}{d\Theta},
	\label{KG_dimensionless}
\end{equation}
where a tilde denotes derivative with respect to $x$, 
and  the dimensionless potential 
\begin{equation}
	V_n(\Theta)=\frac{U_0}{(1+\Theta^2)^n}.
	\label{potential_dimensionless}
\end{equation}
Accordingly, the energy density of the scalar field, normalized by the present
critical density $\rho_{\rm crit}=3H_0^2M_P^2$, reads as
\begin{equation}
	\Omega_\phi = \frac{1}{3}\left[\frac{1}{2}\alpha^2\tilde{\Theta}^2+V_n(\Theta)\right].
	\label{Omega_phi}
\end{equation}


Solving eq.(\ref{KG_dimensionless}) with the explicit values of $U_0 = 3 \times 10^9$, $\alpha = 0.05$, and $\Theta_i = 10^{-3}$ where $\Theta_i=\phi_i/f$ is the initial value of field,  Fig. \ref{background} shows the numerical evolutions of $\rho_\phi(z)$ (top), $V_{n}(\Theta)$ (bottom left) and $\Theta(z)$ (bottom right) for three representative values of $n=\{6, 8,10\}$.
As seen in the top plot, $\Omega_\phi$ is a constant due to the fixed value of $\Theta$ in the high redshift region, 
then becomes of the order of $\sim 10\%$ relative to the total energy density at $z\sim 10^{3.6}$,
and afterward dilutes faster than the radiation energy density in the low redshift region, 
reflecting the nature of this scalar field as an EDE.

Since the late-time residual of EDE is sensitive to the value of $n$,
the range of $n$ favored is quite narrow. 
Given smaller values of $n\leq 6$,  
the scalar field does not fully decay but rises again at $z\lesssim 10$ as inferred from the results of Fig.\ref{background}, indicating an unacceptably large residual energy density 
at the late-time Universe. 
On the other hand, the improvement in the inferred value of $H_0$ becomes marginal for $n>10$. 
As a result, we choose two representative values of $n=\left \{ 8,10 \right \} $ in the following numerical analysis. 

In practical MCMC analysis of the EDE model, one employs the phenomenological parameters, namely $f_{\rm{EDE}}=\rho_{\phi}/\rho_{\rm{tot}}$ and redshift $z_c$ at which the value of $f_{\rm{EDE}}$ is maximal, rather than the model parameters $V_0$ and $f$ in eq.(\ref{eq:potential}).
In appendix.\ref{sa}, we present the details of how to link the model parameters $V_0$ and $f$ to the 
phenomenological parameters $f_{\rm{EDE}}$ and $z_c$ using the shooting algorithm \cite{Smith:2019ihp,Agrawal:2019lmo}.

\subsection{Perturbation evolution}
According to the Klein-Gordon equation in eq.(\ref{eq:KG}), 
the linear perturbation of the scalar $\phi$ in synchronous gauge reads as \cite{Smith:2019ihp,Hill:2020osr},
\begin{eqnarray}
	\delta{\phi}''+2\mathcal{H}\delta {\phi}'+(k^2+a^2V_{,\phi \phi })\delta\phi = -\frac{1}{2}{\phi}'{h}',   
        \label{perturbation}
\end{eqnarray}
where $V_{,\phi \phi }\equiv d^2V(\phi)/d\phi^2$,  a prime denotes a derivative with respect to conformal time, 
and $k$ is the wavenumber.

\section{MCMC analysis}
\label{MCMC}
In this section we carry out the MCMC analysis on the EDE model with two representative values of $n=\{8,10\}$ 
in terms of implementing the background and linear perturbation equations in Sec.\ref{model} into the Boltzmann solver \texttt{CLASS} \cite{Lesgourgues:2011re,Blas:2011rf} \footnote{\url{https://github.com/lesgourg/class_public}} and \texttt{CLASS\_EDE} \cite{Hill:2020osr}\footnote{\url{https://github.com/mwt5345/class_ede}}  interfaced with \texttt{Cobaya} \cite{Torrado:2020dgo,cobaya2019}.
We use \texttt{GetDist} \cite{Lewis:2019xzd} to obtain the posterior distributions and contour plots. 

\subsection{Datasets}
\label{data}
We fit the EDE model to the following datasets:
\begin{itemize}
\item  \textbf{CMB}: we use the Planck 2018 low-\(\ell\) temperature and polarization likelihood \cite{Planck:2019nip},  the high-\(\ell\) temperature, polarization spectra, and the TT, TE, and EE measurements from the NPIPE CamSpec data release \cite{Rosenberg:2022sdy}, and finally  the Planck 2018 CMB lensing likelihood \cite{Planck:2018lbu}.  
\item \textbf{BAO}: the BAO measurements from the DESI DR2 data release \cite{DESI:2025zgx}, which provide both isotropic and anisotropic distance constraints over the redshift interval covered by the survey.
\item \textbf{SN}: the Type Ia supernova distance measurements from the Pantheon+ \cite{Brout:2022vxf,Scolnic:2021amr} compilation.
\item \textbf{H0DN}: we adopt a Gaussian prior on the Hubble constant, $H_0 = 73.50 \pm 0.81$\ km/s/Mpc, based on the baseline results from the $\mathrm{H0DN}$ Collaboration \cite{H0DN:2025lyy}. The Distance Network provides the most precise direct measurement of the Hubble constant to date with an uncertainty of 1.09\%.
\end {itemize}

\subsection{Method} 

\begin{itemize}
\item We impose uniform priors on the EDE parameters as follows:
\begin{equation}
	\begin{aligned}
		\Theta_{\mathrm{i}} &\sim \mathcal{U}(0, 0.1), \\
		f_{\mathrm{EDE}} &\sim \mathcal{U}(0, 0.3), \\
		\log_{10}(z_{c}) &\sim \mathcal{U}(3.0, 5.0);
	\end{aligned}
\end{equation}
 \item The convergence of the chains is checked using the Gelman--Rubin criterion \cite{Gelman:1992zz} with the requirement $R-1 < 0.01$.
\item  We adopt two metrics widely used in the literature to quantitatively measure the goodness of the fit.
The first is the difference in the maximum a posteriori estimator \cite{Raveri:2018wln} defined as
\begin{eqnarray}
	Q_{\mathrm {DMAP}} & = & \sqrt{\chi^2_{min}(\mathcal{D}+\mathrm{H0DN})-\chi^2_{min}(\mathcal{D}) },
\end{eqnarray}
where $\mathcal{D}$ denotes the baseline datasets without the $\mathrm{H0DN}$ prior. 
$Q_{\mathrm {DMAP}}$ can quantify the impact of the local \(H_0\) prior.
The second one is the difference in Akaike information criterion \cite{Akaike:1974vps}:
\begin{eqnarray}
	\Delta \mathrm{AIC} & = & \chi^2_{\rm{min}}(\mathrm{EDE})-\chi ^2_{\rm{min}}(\mathrm{\Lambda CDM})+2\mathcal{N} .
\end{eqnarray}
where $\mathcal{N}=3$ the number of free parameters beyond $\Lambda$CDM.
A negative value of \(\Delta {\rm AIC}\) will indicate a preference of the EDE model over the $\Lambda$CDM for the specific datasets considered. 
\end {itemize}

\subsection{Results}
\label{nr}
We now present the numerical results of the EDE model through the set of parameters $\{\Theta_i, f_{\rm{EDE}},z_{c}\}$ using the shooting algorithm \cite{Smith:2019ihp,Agrawal:2019lmo} as detailed in appendix.\ref{sa}. 
 
\begin{figure}
	\centering
	\includegraphics[width=15cm, height=18cm]{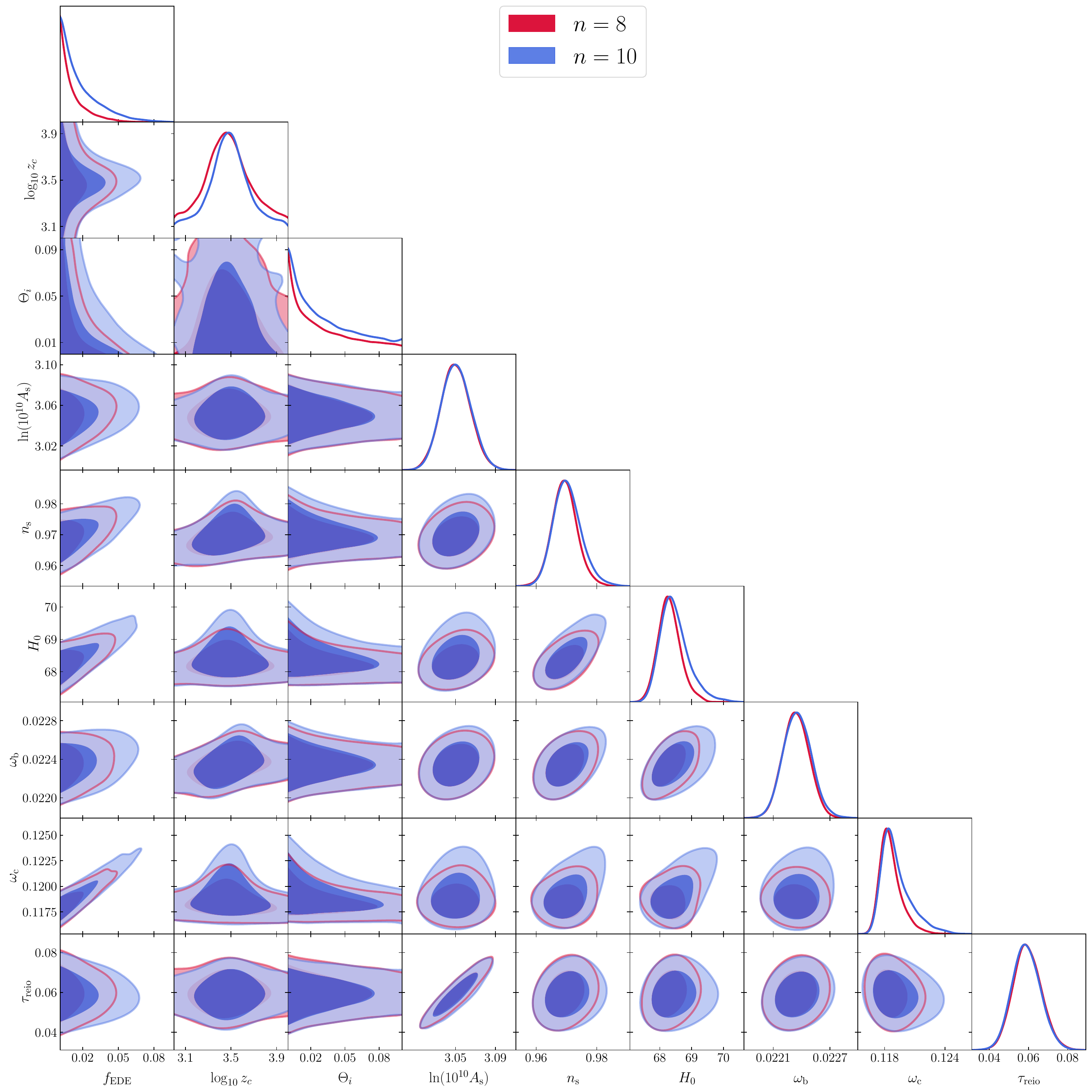}
	\centering
	\caption{Marginalized posterior distributions of the EDE parameters and selected standard cosmological parameters obtained from the datasets of CMB+BAO+SN. The dark and light shaded regions enclose the $68\%$ and $95\%$ CL, respectively.}
	\label{triangle1}
\end{figure}

Table \ref{bestfit} shows the posterior mean values with $1\sigma$ uncertainties and 
the best-fit values (in parentheses) for the EDE model with the values of $n=\{8,10\}$, as compared to the $\Lambda$CDM,
based on the datasets of CMB+BAO+SN.
We find $f_{\rm EDE}<0.0140$ and $f_{\rm EDE}<0.018^{+0.004}_{-0.018}$ for $n=8$ and $n=10$ respectively,
showing no preference of the EDE model over the $\Lambda$CDM as verified by Fig.\ref{triangle1}.
In practice, the positive values of $\Delta{\rm AIC}$ in Table \ref{bestfit} already indicate that the datasets disfavor the additional EDE parameters.

\begin{table}
	\begin{center}
		\centering
		\resizebox{\textwidth}{!}{
			\renewcommand{\arraystretch}{1.5}
			\begin{tabular}{|c|c|c|c|c|}
				\hline\hline
				~ &$\Lambda$CDM & EDE ($n=8$) & EDE ($n=10$) \\ \hline
				
				$f_{EDE}$
				& $--$ 
				& $<0.0140(0.0287)$ 
				& $<0.018(0.030)^{+0.004}_{-0.018}$  \\
				
				$log_{10}z_c$
				& $--$ 
				& $3.48(3.39)^{+0.20}_{-0.24}$ 
				& $3.49(3.47)^{+0.18}_{-0.20}$ \\     
				
				$\Theta_i$ 
				& $--$
				& $<0.0416(0.0010)$ 
				& $<0.0410(0.0019)$ \\   \hline       
				
				$H_{0}$
				& $68.12(68.11)^{+0.30}_{-0.27}$
				& $68.32(68.50)^{+0.31}_{-0.39}$ 
				& $68.47(68.79)^{+0.33}_{-0.51}$ \\
				
				$\omega_{b}$ 
				& $0.02231(0.02229)\pm0.00012$  
				& $0.02234(0.02224)\pm0.00014$
				& $0.02236(0.02237)\pm0.00015$ \\
				
				$\omega_{\rm{c}}$ 
				& $0.11779(0.11785)^{+0.00059}_{-0.00067}$
				& $0.11863(0.12064)^{+0.00067}_{-0.00130}$ 
				& $0.11916(0.12037)^{+0.00082}_{-0.00170}$ \\
				
				$\ln{(10^{10}A_{s})}$ 
				& $3.047(3.040)\pm0.015$
				& $3.050(3.048)\pm0.015$ 
				& $3.051(3.052)\pm0.015$ \\
				
				$n_{s}$ 
				& $0.9677(0.9655)\pm0.0034$ 
				& $0.9695(0.9675)^{+0.0038}_{-0.0043}$ 
				& $0.9704(0.9723)^{+0.0040}_{-0.0049}$ \\
				
				$\tau_{\rm{reio}}$ 
				& $0.0586(0.0566)^{+0.0068}_{-0.0075}$
				& $0.0594(0.0546)\pm0.0073$ 
				& $0.0589(0.0577)\pm0.0072$  \\ 
				
				$S_{\rm{8}}$ 
				& $0.8099(0.807)\pm0.0082$
				& $0.8141(0.8226)^{+0.0083}_{-0.0099}$ 
				& $0.8149(0.8206)^{+0.0087}_{-0.0099}$  \\		\hline
				
				$\Delta\text{AIC}$ 
				& $+0.0$
				& $+6.2$ 
				& $+5.1$  \\

				\hline \hline
			\end{tabular}
		}
		\caption{Posterior mean values and $1\sigma$ uncertainties of the EDE model parameters compared to the $\Lambda$CDM based on the datasets of CMB+BAO+SN. The best-fit values are given in parentheses.}
		\label{bestfit}
	\end{center}
\end{table}

\begin{table}
	\begin{center}
		\centering
		\resizebox{\textwidth}{!}{
			\renewcommand{\arraystretch}{1.5}
			\begin{tabular}{|c|c|c|c|c|}
				\hline\hline
				~ &$\Lambda$CDM & EDE ($n=8$) & EDE ($n=10$) \\ \hline
				
				$f_{EDE}$
				& $--$ 
				& $0.067(0.075)\pm0.021$ 
				& $0.079(0.091)^{+0.024}_{-0.020}$  \\
				
				$log_{10}z_c$
				& $--$ 
				& $3.471(3.410)^{+0.080}_{-0.100}$ 
				& $3.499(3.552)^{+0.084}_{-0.075}$ \\     
				
				$\Theta_i$ 
				& $--$
				& $<0.0038(0.0003)$ 
				& $0.0098(0.0011)^{+0.0015}_{-0.0097}$ \\   \hline       
				
				$H_{0}$
				& $68.62(68.64)^{+0.33}_{-0.22}$
				& $69.91(70.20)\pm0.49$ 
				& $70.20(70.56)^{+0.60}_{-0.47}$ \\
				
				$\omega_{b}$ 
				& $0.02243(0.02248)^{+0.00013}_{-0.00010}$  
				& $0.02252(0.02224)^{+0.00019}_{-0.00024}$ 
				& $0.02261(0.02264)\pm0.00023$ \\
				
				$\omega_{\rm{c}}$ 
				& $0.1168(0.1169)^{+0.0005}_{-0.0007}$
				& $0.1223(0.1235)\pm0.0019$ 
				& $0.1237(0.1244)\pm0.0021$ \\
				
				$\ln{(10^{10}A_{s})}$ 
				& $3.052(3.054)\pm0.015$
				& $3.063(3.057)\pm0.016$ 
				& $3.066(3.063)\pm0.016$ \\
				
				$n_{s}$ 
				& $0.9704(0.9689)^{+0.0037}_{-0.0032}$ 
				& $0.9796(0.9753)^{+0.0055}_{-0.0062}$ 
				& $0.9825(0.9857)\pm0.0065$ \\
				
				$\tau_{\rm{reio}}$ 
				& $0.0621(0.0636)^{+0.0071}_{-0.0079}$
				& $0.0603(0.0547)^{+0.0071}_{-0.0081}$ 
				& $0.0600(0.0565)^{+0.0070}_{-0.0080}$  \\ 
				
				$S_{\rm{8}}$ 
				& $0.801(0.801)^{+0.007}_{-0.009}$
				& $0.824(0.828)\pm0.012$ 
				& $0.828(0.831)\pm0.012$  \\	 \hline
				
				$\Delta\text{AIC(+H0DN)}$ 
				& $+0.0$
				& $-8.6$ 
				& $-11.1$  \\		\hline
				
				$Q_{\text{DMAP}}$ 
				& 6.25$\sigma$
				& 4.93$\sigma$
				& 4.79$\sigma$  \\				
				
				\hline \hline
			\end{tabular}
		}
		\caption{The same as in Table \ref{bestfit} but with the datasets of CMB+BAO+SN+H0DN instead.}
		\label{bestfit2}
	\end{center}
\end{table}

\begin{figure}
	\centering
	\includegraphics[width=15cm, height=18cm]{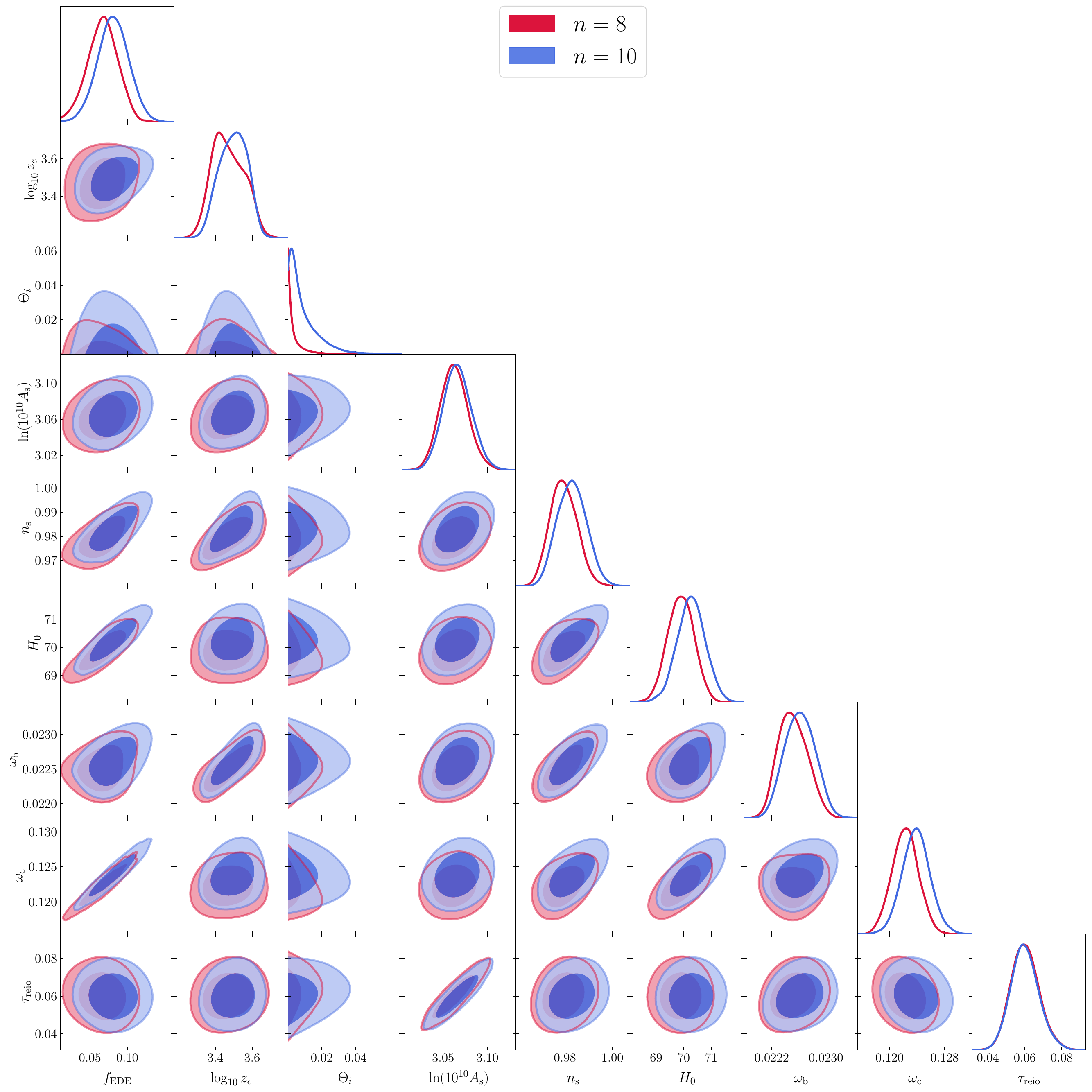}
	\centering
	\caption{The same as in Fig.\ref{triangle1} but with the datasets of CMB+BAO+SN+H0DN instead.}
	\label{triangle2}
\end{figure}

The situation changes after including the H0DN prior, as shown in Table \ref{bestfit2} and Fig.\ref{triangle2}. 
We obtain $f_{\rm EDE}=0.067\pm0.021$ and $f_{\rm EDE}=0.079^{+0.024}_{-0.020}$ for $n=8$ and $n=10$, respectively.
For $n=8$ one finds that $\Delta{\rm AIC}=-8.6$, $H_0=69.91\pm0.49\,{\rm km\,s^{-1}\,Mpc^{-1}}$ and $Q_{\text{DMAP}}=4.93\sigma$.
For $n=10$,  $\Delta{\rm AIC}=-11.1$,  the inferred $H_0$ is higher, and $Q_{\text{DMAP}}$ decreases mildly to 4.79$\sigma$.
The values of $\Delta{\rm AIC}$ favor the preference of EDE over the $\Lambda$CDM for the datasets of  CMB+BAO+SN+H0DN, 
whereas the values of $Q_{\text{DMAP}}$ suggest that the conflict between the same datasets and the $\Lambda$CDM becomes weaker in the EDE model as a result of the inclusion of the H0DN likelihood. 

\begin{figure}
\centering
\includegraphics[width=8cm,height=8cm]{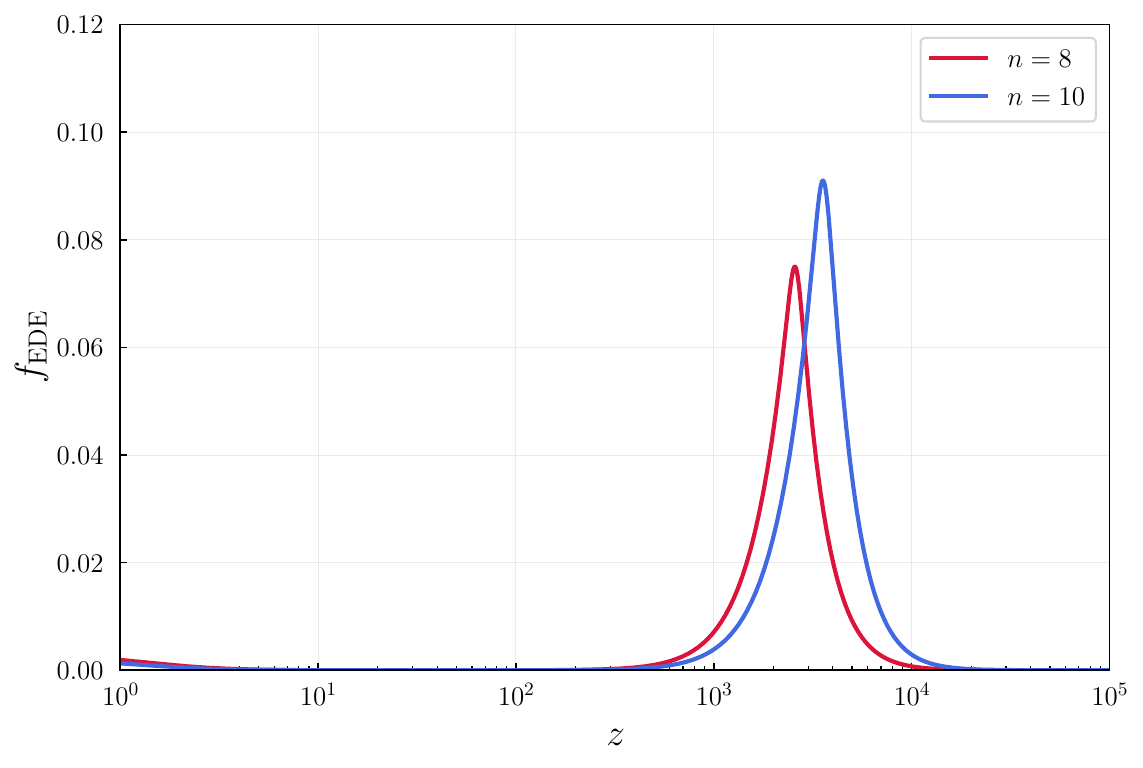}
\includegraphics[width=8cm,height=8cm]{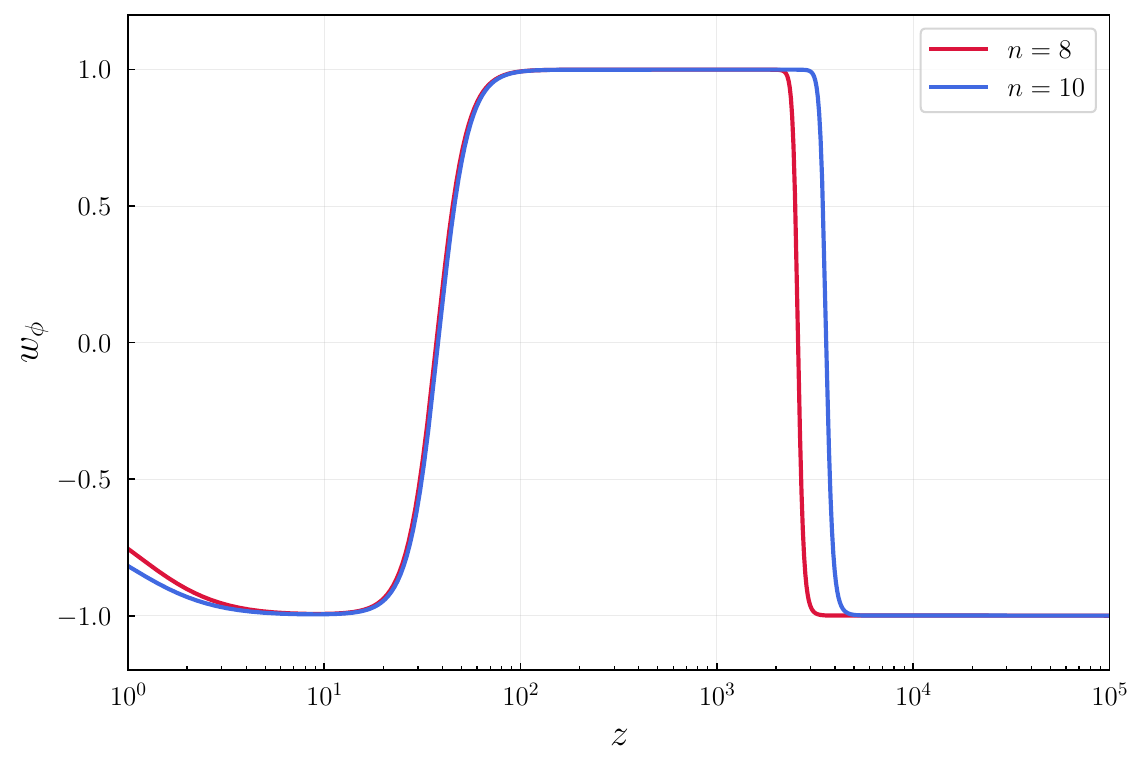}
\centering
\caption{The left (right) shows the evolution of $f_{\rm EDE}$ ($w_{\phi}$) as function of the redshift for different values of $n=\{8,10\}$, using the best-fit values of the EDE model and cosmological parameters in Table \ref{bestfit2}.}
\label{evolution}
\end{figure}

Fig.\ref{evolution} shows the evolutions of $f_{\rm EDE}$ and $w_{\phi}$ as a function of redshift $z$ using the best-fit values of the EDE model and cosmological parameters in Table \ref{bestfit2}. 
As seen in the left panel, the EDE fraction of energy density indeed peaks around $\log_{10}z_c\simeq3.5$, close to matter--radiation equality, which rapidly decreases at the late-time Universe. 
However, a small residual remains at $z\sim 0$, being significant for $n<8$ as mentioned in Sec.\ref{model}.

\section{Detectable signals}
\label{signals}

The numerical results in Table \ref{bestfit2} show that our EDE model may leave observable signals in the CMB and matter power spectra.

\begin{figure}
	\centering
	\includegraphics[width=15cm, height=16cm]{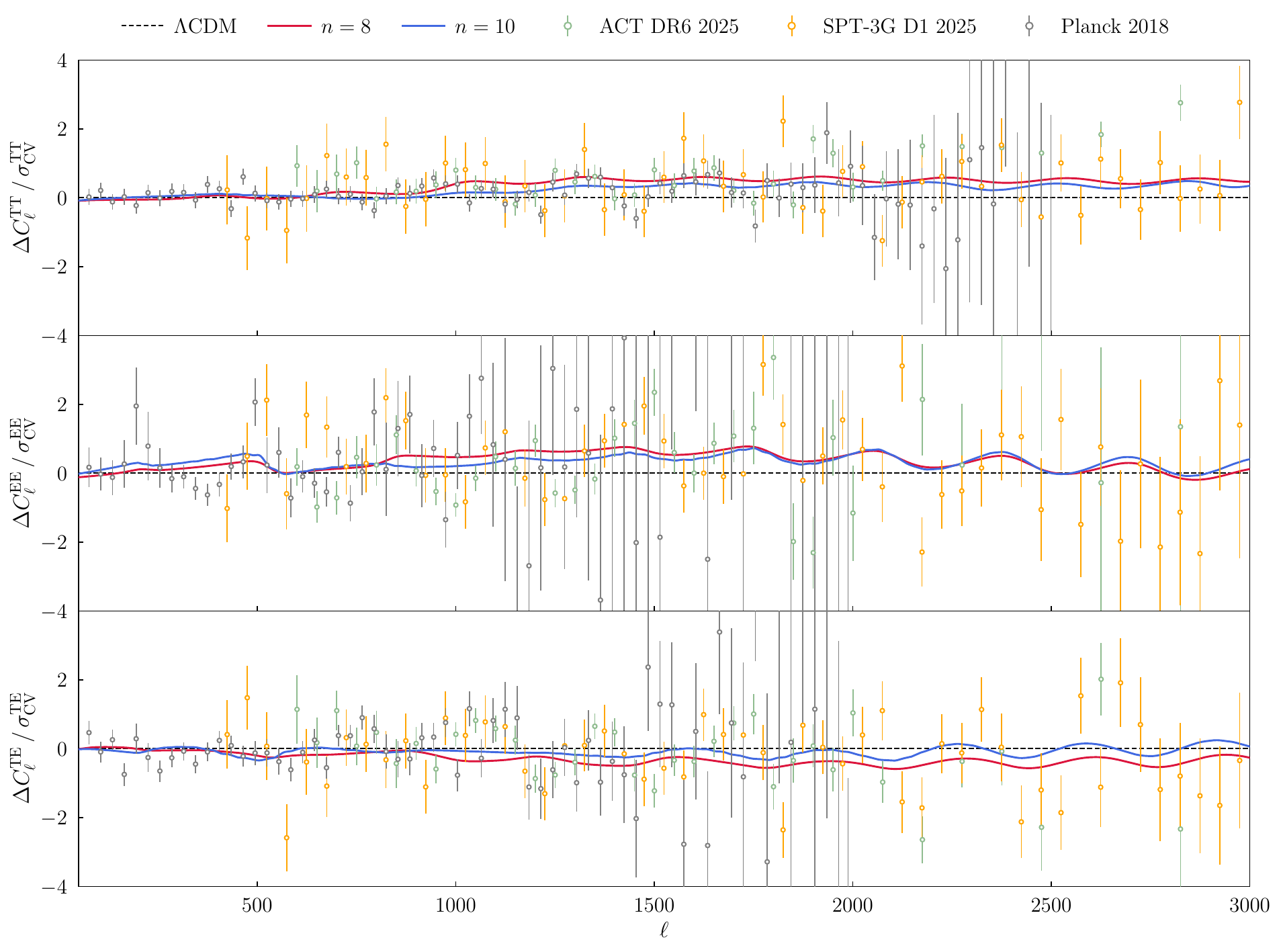}
	\centering
	\caption{Deviations in the CMB temperature and polarization power spectra relative to the $\Lambda$CDM normalized by the CV limits, which are compared to the Planck 2018 \cite{Planck:2018vyg}, ACT DR6 \cite{AtacamaCosmologyTelescope:2025blo}, and SPT-3G D1 \cite{SPT-3G:2025bzu} data points.
Here, we have used the best-fit values of parameters in Table \ref{bestfit2} obtained from the datasets of CMB+BAO+SN+H0DN,
the red and blue curves correspond to $n=8$ and $n=10$ respectively, and the dashed black line denotes the $\Lambda$CDM reference.}
	\label{CMB}
\end{figure}

\subsection{CMB power spectra}
\label{cmbps}

Adding the EDE results in shifts in the cosmological parameters as seen in Table \ref{bestfit2}, 
in order to maintain consistency with the CMB data \cite{Poulin:2023lkg, Hill:2020osr}.
At the background level, the EDE reduces the angular sound horizon and damping scales compared to the $\Lambda$CDM,  leading to a higher TT and EE power spectra at large $\ell$. 
At the perturbative level, it changes the Weyl potential, modifying the CMB modes entering the horizon right around $z_c$, corresponding to a reference value of $\ell_{c}\sim 100$ for $z_{c}\sim 10^{3.5}$.
To see these effects clearly, we present in Fig.\ref{CMB}  the deviations in the CMB temperature and polarization power spectra relative to the $\Lambda$CDM normalized by the cosmic variance (CV) limits, which are compared to the Planck 2018 \cite{Planck:2018vyg}, ACT DR6 \cite{AtacamaCosmologyTelescope:2025blo}, and SPT-3G D1 \cite{SPT-3G:2025bzu} data points.
Therein we have used the best-fit values of the EDE and cosmological parameters in Table \ref{bestfit2} obtained from the datasets of CMB+BAO+SN+H0DN.
This figure shows that the deviations remain well below one cosmic variance unit over the multipole range displayed,
and may be detectable with future CMB data.

\begin{figure}
	\centering
	\includegraphics[width=16cm, height=8cm]{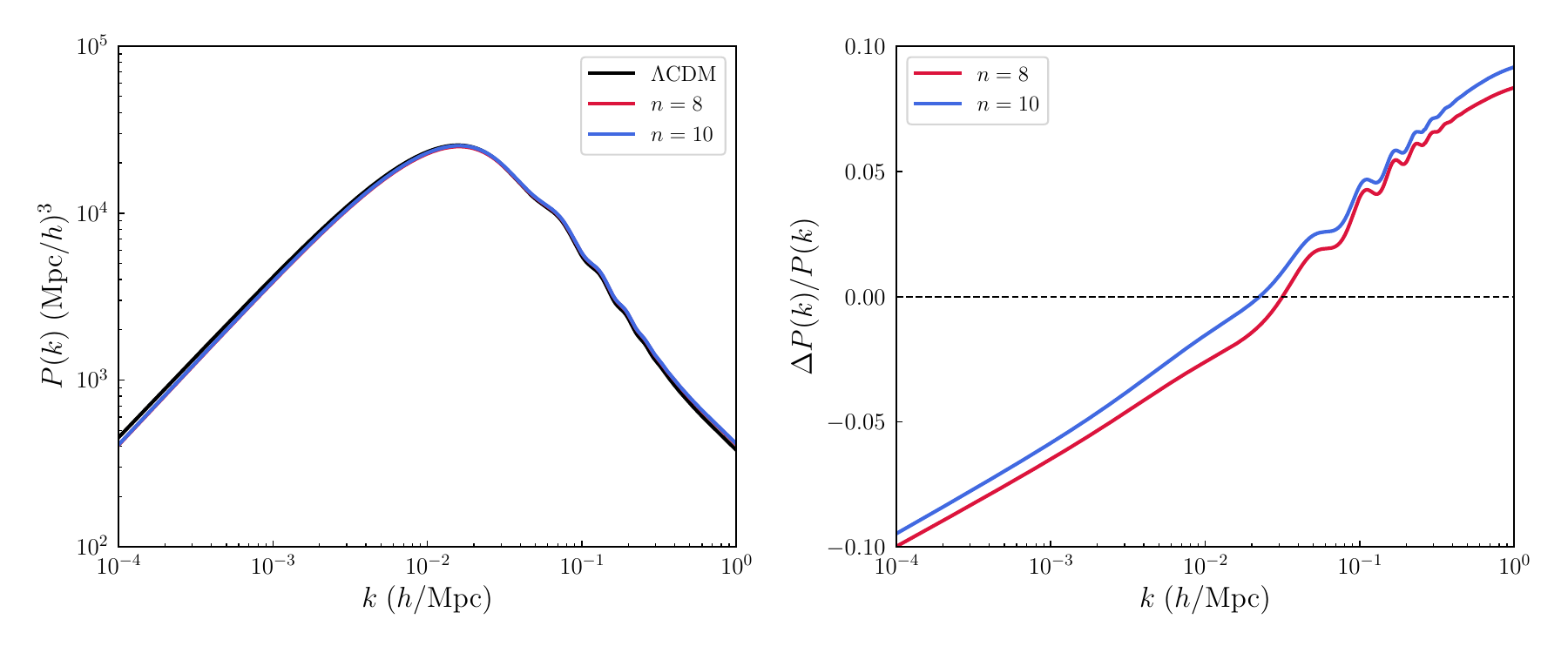}
	\centering
	\caption{Matter power spectra $P(k)$ relative to the $\Lambda$CDM, using the best-fit values of  the EDE and cosmological parameters in Table.\ref{bestfit2}  obtained from the datasets of CMB+BAO+SN+H0DN. The left panel shows $P(k)$, with black, red, and blue curves denoting $\Lambda$CDM, EDE with $n=8$ and $n=10$ respectively. The right panel shows $\Delta P(k)/P_{\Lambda{\rm CDM}}(k)$, with $\Delta P(k)=P_{\rm EDE}(k)-P_{\Lambda{\rm CDM}}(k)$. }
	\label{pk}
\end{figure}

\subsection{Matter power spectra}
\label{mps}
At the background level, the EDE results in an increase of the DM density and primordial tilt compared to the $\Lambda$CDM, 
which tend to worsen the $S_8$ tension as previously emphasized in \cite{Poulin:2023lkg}. 
To see the effects on the matter power spectra at lower $k$, 
we show in Fig.\ref{pk} the deviations in $P(k)$ relative to the $\Lambda$CDM over the range of $k\sim 10^{-4}-1$ h/Mpc,
where we have used the best-fit values of the EDE and cosmological parameters in Table \ref{bestfit2} obtained from datasets of CMB+BAO+SN+H0DN.
Fig.\ref{pk} shows that the deviations are $k$-dependent, reaching  $\sim 10\%$ near $k\sim1$ h/Mpc.

Note that, there are already large-scale  data about $\Delta P(k)$ over the $k$ range in Fig.\ref{pk} , see \cite{Chabanier:2019eai} and references therein.
Nevertheless,  these data points are highly model-dependent as mentioned in \cite{Chabanier:2019eai}, 
which cannot be directly used to place constraints on the EDE parameters.
To extract the large-scale constraints, one has to include the EDE model at the beginning of data analysis, 
as in the study of BOSS constraints on the axion-like EDE \cite{Ivanov:2020ril}.
This topic is beyond the scope of the present work.

\newpage
\section{Conclusion}
\label{con}
In this work we have proposed a new EDE model to alleviate the Hubble tension.
The EDE is a scalar field $\phi$ with the potential that has an inverse power-law asymptotic tail, 
which is different from those of the existing EDE models in the literature. 
The main features are 
\begin{itemize}
\item From the perspective of phenomenology, the datasets of CMB+BAO+SN+H0DN prefer this EDE model over the $\Lambda$CDM, 
with the  68$\%$ CL value of $H_0\sim 70.20^{+0.60}_{-0.47}$ km s$^{-1}$Mpc$^{-1}$ 
and the best-fit value of $H_0\sim 70.56$ km s$^{-1}$Mpc$^{-1}$ for the index $n=10$.
Moreover, the value of $S_8$ also increases as in the other EDE models, worsening the $S_8$ tension.
Finally,  the EDE model can be tested by future data of CMB and matter power spectra.
\item From the theoretical perspective, the EDE model suffers from the fine-tuning issues.
The first one is the initial condition of $\Theta_{i}\leq 10^{-3}$.
Second, it is not clear whether such kind of potentials with an inverse power-law asymptotic tail can be embedded into ultraviolet complete theory.
\end{itemize}

Following our study, there are a few interesting points left for future investigations. 
On one hand, the tension level can be further reduced in alternative  inverse power-law EDE models such as 
$V(\phi)=V_0 \rm{sech}^{n}(\phi/f)$ which has a similar shape of potential but with a steeper slope.
On the other hand, including additional datasets, especially galaxy clustering data, may affect the main numerical results presented in this work.

\section*{Acknowledgements}
We acknowledge the use of \texttt{CLASS} \cite{Lesgourgues:2011re,Blas:2011rf}, \texttt{CLASS\_EDE} \cite{Hill:2020osr}, \texttt{Cobaya} \cite{Torrado:2020dgo,cobaya2019} and \texttt{Getdist} \cite{Lewis:2019xzd}.
The codes are available from the authors upon reasonable request.

\appendix 
\setcounter{equation}{0}\renewcommand\theequation{A\arabic{equation}}

\section{Shooting algorithm}
\label{sa}
We introduce the shooting algorithm \cite{Smith:2019ihp,Agrawal:2019lmo} - a method of linking the model parameters $V_0$ and $f$ to phenomenological parameters $f_{\rm{EDE}}$ and $z_c$ used by the MCMC analysis.

Given a set of input values $z_c$ and $f_{\rm{EDE}}$ in the Boltzmann code, 
the shooting algorithm finds the initial guess values of  $V_0$ and $f$ via 
\begin{eqnarray}
V_0\approx \frac{3f_{\rm{EDE}}}{1-f_{\rm{EDE}}}M^{2}_{P}H^{2}_{0}E^{2}(1+\Theta^{2}_i)^{n},
\end{eqnarray}
and 
\begin{eqnarray}
V_{,\phi\phi}(\Theta_i)=-\frac{2nV_0}{f^{2}}\frac{1-(2n+1)\Theta^{2}_i}{(1+\Theta^{2}_i)^{n+2}},
\end{eqnarray}
where the values of $n$ and $\Theta_i$ are known.
The value of $\Theta$ is initially frozen as $\Theta\approx \Theta_i $ at high $z$ regions, 
then begins to roll around $z_c$ at which $\mid V_{,\phi\phi}(\Theta_i)\mid \sim \kappa H^{2}(z_{c})$, with coefficient $\kappa$ determined numerically. 
Plugging the initial guess values of  $V_0$ and $f$ into the background equations, 
one obtains the derived values of $z_c$ and $f_{\rm{EDE}}$.
If the deviations between the input and derived values  of $z_c$ and $f_{\rm{EDE}}$ are acceptable,
the link is established.
Otherwise, we perform an iterate process of altering $V_0$ and $f$ from their guess values to find the new derived values of $z_c$ and $f_{\rm{EDE}}$ until the required precisions are achieved.

\end{document}